# PANSPERMIA, PAST AND PRESENT:

## Astrophysical and Biophysical Conditions for the Dissemination of Life in Space.

Paul S. Wesson[1,2]

[1]Department of Physics and Astronomy, University of Waterloo, Waterloo, Ontario
 N2L 3G1, Canada

[2]Herzberg Institute of Astrophysics, Victoria, B.C.  V9E 2E7, Canada

Abstract:    Astronomically, there are viable mechanisms for distributing organic material throughout the Milky Way.  Biologically, the destructive effects of ultraviolet light and cosmic rays means that the majority of organisms arrive broken and dead on a new world.  The likelihood of conventional forms of panspermia must therefore be considered low.  However, the *information* content of damaged biological molecules might serve to seed new life (necropanspermia).

Correspondence: Mail to address 2 above; email = psw.papers@yahoo.ca

# PANSPERMIA, PAST AND PRESENT

1. Introduction

Panspermia is underlain by the idea that the vast number of stars in the Milky Way is somehow matched by the fecundity of life. Astronomical conditions are such that it appears feasible to transport organic material from one star system to another; but those same conditions mean that biological damage is severe (notably due to ultraviolet light and cosmic rays), so the majority of organisms arrive on a new world in an inactivated or dead state. The majority opinion is that while organisms may be ejected from an Earth-like planet by the collision of an asteroid or comet, the DNA and RNA is so degraded in space that the probability of seeding life in the Galaxy is low. Despite this, interest in the origin of life is high, and mechanisms continue to be investigated that might lead to a viable version of panspermia.

Lithopanspermia, wherein life is distributed through space on rocks such as meteorites, was suggested by Lord Kelvin in an address to the British Society for the Advancement of Science in 1871. Radiopanspermia, which involves the movement of organisms by the radiation pressure of the Sun and other stars, was discussed by Arrhenius in his book *Worlds in the Making* of 1908. Subsequent to these two groundbreaking suggestions, the possible role of cosmic rays was discussed, since the low-energy particles which form the solar wind and the fluxes of ionizing radiation from other stars join with higher-energy particles from astrophysical sources to inactivate (or kill) micro-organisms in space. Directed panspermia was proposed by Crick and Orgel (1973) in an attempt to circumvent both the transport and inactivation problems, and envisages a technologically-advanced alien civilization which purposefully sent spaceships laden with micro-organisms to seed life throughout the early Milky Way. Recently, however, it has been argued that the same end may be achieved by natural mechanisms under favourable circumstances (Wallis and Wickramasinghe 2004). Microbial life can be protected from ultraviolet photons and cosmic-ray particles if it is buried in rocky objects of 1 mm - 1 m size, and after traversing the comet-laden outer parts of a planetary system and being slowed by a protostellar disk or giant molecular cloud (Napier and Staniucha 1982), the surviving organisms may be deposited on a remote planet, which if hospitable would provide a new home for life.

Variations on these versions of panspermia are numerous (Hart 1982; Cosmovici, Bowyer and Wertheimer 1997; Hoyle and Wickramasinghe 2000; Burchell 2004; Sullivan and Baross 2007). Here, what may be termed the standard model is assumed. In this, micro-organisms in rocky material are ejected into space following an impact by an asteroid or comet on an Earth-like planet, the material being ejected from the solar system, to travel through interstellar space before entering another planetary system, where the organisms are deposited on a new world. Detailed calculations on this model have been done by several groups (Secker, Lepock and Wesson 1994; Clark et al. 1999; Mileikowsky et al. 2000; Melosh 2003; Valtonen et al. 2009). The consensus view is that radiation pressure is the dominant mechanism for accelerating organism-laden material of grain size out of the first planetary system and decelerating it into the second system; and that ultraviolet light is the dominant mechanism for inactivating organisms in space, which even if they are partially shielded by rocky material happens on a timescale of $10^6$ yr (see below). However, there are significant differences of opinion about the contending influences of electromagnetic radiation and cosmic rays. In Sections 2 and 3 below, the astrophysical and biophysical conditions around panspermia will be examined anew. A simplified, bolometric calculation will be done which complements an earlier spectral one (Secker, Lepock and Wesson 1994). This will confirm that radiation pressure is the dominant transport mechanism, and that most micro-organisms are indeed killed by ultraviolet light. However, it will be noted that because of their sizes and nature, a possible niche for panspermia might be filled by viruses or fragments of them. In fact, a new aspect of panspermia is revealed if we concentrate not on the question of alive-versus-dead (which is difficult to evaluate), but on the question of genetic *information* (which can be quantified; Wesson, Secker and Lepock 1997). Since viruses are commonly regarded as intermediate between 'living' and 'dead', and are not normally considered as panspermia agents, a short account of their relevant properties is given in the Appendix. Section 4 is a conclusion.

It will become clear that considerable uncertainties remain about panspermia, even though significant advances have been made recently. Of these, a major one is the realization that a few meteorites found on Earth must have originated on the Moon and on the planet Mars. This confirms the feasibility of the first stage of panspermia, as argued by Melosh (1988, 2003), though he and other workers have voiced skepticism about the later stages of the process. Another discovery is that some forms of life – the extremophiles – exist on the Earth in environments that are very inhospitable compared to what was previously considered normal. Some of the uncertainty about panspermia is related to another question that is notoriously troublesome, namely how to define 'life' (Cleland and Chyba 2007; Shapiro 2007; Abbot, Davies and Pati 2008). This topic is beyond the scope of the present account. But the overall conclusion to be drawn is that while astrophysical aspects of the theory are quite firm, the biophysical aspects are unsure, and are the main subjects that need further study before anything definitive can be stated about the viability of panspermia.

2. Astrophysical Aspects of Panspermia

The aim of this section is to clarify the status of electromagnetic radiation (mainly ultraviolet) and particle radiation (mainly the solar wind and other cosmic rays) in the transport and inactivation of micro-organisms in dust grains in space. The primary quantity to be calculated is the ejection speed from a planetary system such as ours, which governs the time of transport to another one in the Milky Way.

Consider a star like the Sun with luminosity $L$, whose flux at radius $R$ is $L/4\pi R^2$. The energy density of the photons at $R$ is $L/4\pi c R^2$ where $c$ is the speed of light. Each photon has energy $e = hc/\lambda$ where $h$ is Planck's constant and $\lambda$ is the wavelength. Thus the number density of photons is $L\lambda/4\pi R^2 c^2 h$. The number of photons which impact a grain in a second is the last quantity times $c$ times the grain cross-section $\sigma$. Each impacting photon carries a momentum $p = e/c = h/\lambda$, so the momentum flux is the product of these quantities:

$$dp/dt = L\sigma/4\pi R^2 c \quad . \tag{1}$$

This photon momentum flux, assuming it is transformed efficiently to the grain, causes the latter to feel a force $m\,dv/dt$ where $m$ and $v$ are the grain's mass and velocity. However, $v = dR/dt$ in terms of the radial distance travelled in time $t$, so $dt = dR/v$. We can then combine relations and integrate, to obtain the velocity $v = v(R)$ as a function of distance from the central source: $v^2 = (2L\sigma/4\pi cm)\int R^{-2}dR$. This shows that most of the impetus to the grain comes from the inner region, where the radiation is more intense. We can to a good approximation therefore neglect the contribution to the noted integral from the outer limit as compared to the inner one, $R_o$ (say). Also, it is convenient to replace $\sigma$ and $m$ by the radius $r$ and density $\rho$ of the grain, which we take to be spherical and uniform, so $\sigma = \pi r^2$ and $m = (4/3)\pi r^3 \rho$. Then the velocity of the grain is

$$v = \left(\frac{3L}{8\pi cr\rho R_o}\right)^{1/2} . \qquad (2)$$

In this, we can insert the luminosity of the Sun ($4 \times 10^{33}$ erg s$^{-1}$) and measure the starting radius of the grain in A.U. ($1.5 \times 10^{13}$ cm). Numerically then, to a good approximation,

$$v = 3.3 \times 10^4 (r\rho R_0)^{-1/2} . \qquad (3)$$

For $r = 1 \times 10^{-5}$ cm, $\rho = 3$ g cm$^{-3}$ and $R_o = 1$ A.U., this gives $v \simeq 6 \times 10^6$ cm s$^{-1}$. That is, a silicate grain in space at the distance of the Earth's orbit from the Sun will be pushed outwards in the solar system by radiation pressure at a speed of approximately 60 km s$^{-1}$.

This is slightly larger than the gravitational escape velocity from the solar system of an object near the Earth's orbit, which is approximately 45 km s$^{-1}$. That some kind of balance between radiation pressure and gravitation may exist is reasonable, given that both influences fall off with distance as $1/R^2$. Such a balance can be expressed by equating the radiation force (1) to the gravitational force between the Sun (mass $M$) and the grain (mass $m$):

$$\frac{L\sigma}{4\pi R^2 c} = \frac{GMm}{R^2}. \tag{4}$$

This is most instructively presented in terms of the radius ($r$) and density ($\rho$) of the grain, so in general and numerically we have

$$r\rho = \frac{3L}{16\pi cGM} \simeq 5.7 \times 10^{-5}, \tag{5}$$

where $r$ is in cm and $\rho$ is in g cm$^{-3}$. For a silicate grain ($\rho = 3$ g cm$^{-3}$), this gives approximately $r = 2 \times 10^{-5}$ cm. For grains larger than this, gravity is dominant; while for grains smaller than this, radiation is dominant. This can be appreciated from previous relations, which show that the acceleration of a dust grain due to radiation pressure varies with $\sigma/m \sim r^2/r^3 \sim 1/r$. Above, we considered a grain with $r = 1 \times 10^{-5}$ cm, and calculated its radiation-induced velocity as 60 km s$^{-1}$. We infer that grains of this size and smaller will indeed be ejected from the solar system. However, the countervailing effects of gravity mean that we should picture the process as one wherein grains with $r \leq 1 \times 10^{-5}$ cm 'leak' out of the solar system with speeds of order 10 km s$^{-1}$.

The preceding calculation concerned radiation pressure as an ejection mechanism for grains in our solar system. It made certain plausible simplifications. For example, it used an average wavelength for the Sun's radiation, and this led to the input of the total luminosity in (5), so the calculation was effectively a bolometric one. Also, the calcula-tion neglected the contribution from the solar wind and low-energy cosmic rays which originate in the Sun. However, it is easy to show that the latter are ineffective in relative terms. For example, the solar wind at the distance of the Earth's orbit has a number density in the range 1-10 particles (mainly protons) per cubic centimetre, and a speed of 400-600 km s$^{-1}$, with an effective temperature of around 2 x 10$^5$ K. Following a method analogous to that above, the acceleration of a 10$^{-5}$ cm grain at the distance of the Earth's orbit can be calculated. It is of order 1000 times *less* than the acceleration due to radiation pressure (which is of order 1 cm s$^{-2}$). The effect of cosmic rays is likewise negligible by comparison. Of course, it is possible to consider stars of other types and in other phases, including the early T-Tauri phase and the late red-giant phase (Secker, Lepock and Wesson 1994, Overduin and Wesson 2008). However, the majority of stars like the Sun spend most of their lives on the main sequence; and stellar evolution is well understood, as now is the development of the integrated light of stars (Wesson 1991). It is therefore reasonable to take our solar system at the present time as the best source of data. At present, the only significant uncertainty about the Sun's interaction with dust grains concerns the possibility that those with sizes 10$^{-6}$ cm or smaller may be electrically charged and so coupled to the solar wind, in which case their speed of ejection would be closer to 600 km s$^{-1}$ rather than 10 km s$^{-1}$. There is some evidence of this from the STEREO, ACE and WIND spacecraft (Meyer-Vernet et al. 2009; Russell et al. 2010). However, these observations may refer to a localized disturbance, and theory implies that the effects of electrical charge decreases when the size of the grain increases (Spitzer 1968; Wesson 1973). In future, we can expect more data from ground-based observations of extrasolar planetary systems (Marcy et al. 2005) and satellite surveys of remote star fields by projects such as KEPLER (Basri, Borucki and Koch 2005). For now, though, we adopt the conservative estimate that grains with sizes $10^{-5}$cm are ejected into interstellar space with speeds of order 10 km s$^{-1}$.

The inactivation or killing of micro-organisms depends on a host of factors and is therefore controversial. The main issues will be discussed in Section 3 below. For now, we note that both electromagnetic radiation and the particles of cosmic rays can break the relatively weak chemical bonds which are responsible for the integrity and genetic properties of molecules such as DNA and RNA. Unshielded micro-organisms are *promptly* killed by the ultraviolet component of the Sun's light when they find themselves in space, and the majority of even shielded organisms are inactivated in this way after $10^5$ - $10^6$ yr in space (Secker, Lepock and Wesson 1994). Cosmic rays have a deleterious effect on micro-organisms which, depending on their nature, also has a timescale of $10^5$ - $10^6$ yr (Horneck 1982). Observational data are firmer for electromagnetic radiation than cosmic rays. The latter consist of low-energy particles from the Sun (synonymous in part with the solar wind), medium-energy particles from other sources in the Milky Way, and high-energy particles believed to originate in extragalactic sources. The acceleration mechanisms for these are poorly understood, especially for the high-energy component (which however has a relatively low number density). As regards the inactivation of micro-organisms, the consensus view appears to be that electromagnetic radiation is dominant. Thus most practical work has focussed on the effects of short-wavelength light (Horneck 1981, 1993; Greenberg, Weber and Schutte 1984; Weber and Greenberg 1985; Mennig-man 1989; Horneck, Bucker and Reitz 1994; Noetzel et al. 2007).  However, the destructive effects of cosmic rays have been emphasized in the theoretical calculations of Valtonen et al. (2009), following earlier work by Mileikowsky et al. (2000), and this view has been repeated by Burchell and Dartnell (2009).  It may be useful to clarify the situation, using the firm data on the solar radiation and solar wind noted above.  There, it was shown that the radiation dominates the wind concerning the acceleration of a dust grain in space at the Earth's orbit.  A similar calculation can be done concerning the energy absorbed by a similar grain.  Using the same model, it is straightforward to show that the rate at which energy is absorbed by the grain is approximately $\sigma u e n$.  Here $\sigma$ is the cross-section of the grain while $u$, $e$ and $n$ denote the velocity, energy and number density of the impacting particles.  These are either photons or protons, and since $\sigma$ is the same in either case the difference depends on the quantity $u\,e\,n$ which has the physical dimensions of $MT^{-3}$.  For the aforementioned data, this quantity in units of $g\,s^{-3}$ has the sizes $1.4 \times 10^6$ and $0.4$ approximately, for photons and protons respectively.  This result can be adjusted to take into account higher-energy cosmic rays, but since the spectrum falls off steeply the conclusion stands.  That is, not only the acceleration, but also the rate at which energy is absorbed, is dominated by electromagnetic radiation rather than particle radiation.

The only feasible way to avoid this conclusion is to step away from the 'standard' version of panspermia, and argue that organisms are transported not in dust grains but in large boulders.  This is possible in principle, and can work in practice inside the solar system (see below); but is disfavoured for transport over larger distances by statistical arguments (Horneck et al. 2001; Wallis and Wickramasinghe 2004; Adams and Spergel 2005; Warmflash and Weiss 2006). The chances are best when boulders are ejected from the surfaces of planets in orbit around stars which are themselves in some kind of populous cluster. It is, of course, well known that many stars are members of populations from 2 (binaries) to order $10^6$ (globular clusters). For a given cluster, Monte Carlo simulations can be carried out to study the effects of the dynamical variables and to estimate the probability of a life-bearing rock eventually landing on a new planet (Adams and Spergel 2005). For boulders of mass greater than 10kg, one stellar system may eject of order $10^{16}$ such, but only 1 in $10^4$ of these impact other planets, and for a given cluster of stars the chance of a successful panspermia event is in the range $10^{-3}$-1. This means that most of the systems in a given cluster remain unseeded. The probability of life being carried to *other* clusters, or across the Milky Way, is accordingly very small indeed. Thus while the transport of living organisms inside boulders may be viable within a solar system like ours, it is unlikely in a Galactic context, and is disfavoured compared to the traditional version of panspermia with radiation-driven dust grains.

The results derived in this section make it clear that panspermia in its 'standard' version is a light-dominated process, as first argued by Arrhenius (1908). However, it is by no means obvious that the theory is viable in its original form. At a speed of 10 km $s^{-1}$, the time taken to traverse 1 pc ($\simeq$ 3 lightyears) is approximately $10^5$ yr. So nearby stars can be reached in a time $10^5$ - $10^6$ yr. And to travel 10 kpc, the radius of the optical disk of the Milky Way, takes of order $10^9$ yr. This is shorter than the estimated age of about 13 x $10^9$ yr, so in principle micro-organisms can be dispersed throughout the $10^{10}$ - $10^{11}$ star systems in the Galaxy. Unfortunately, depending on how effectively the organisms are shielded, their inactivation timescale is in the range 0 (unshielded) to $10^7$ yr. (well shielded). On a statistical basis, panspermia with living organisms must therefore be regarded as somewhat unlikely.

3. Biophysical Aspects of Panspermia

The aim of this section is to constrain the nature of the ancient biological material which was able to evolve into the life forms we are familiar with today on the Earth. This task may appear, at first, to be difficult. Evolution over billions of years has certainly transformed whatever constituted the precursor of present life; but we will argue that some progress can be made using an *informational* approach. This has certain advantages. It is quantitative, and sidesteps the tricky question of how we define 'life'. Also, it includes the traditional kind of panspermia, as well as the form we are obliged to consider by the results of the preceding section, namely that organisms arrive at a new planet in a broken or dead state.

It has been known for a long while that random chemical interactions cannot produce the genetic information of the organisms we currently see on Earth (Argyle 1977, Hoyle 1980, Hart 1982, Barrow and Tipler 1986, Wesson 1990). For example, in a model of the prebiotic Earth with an appropriate complement of amino acids, random molecular interactions over a period of $500 \times 10^6$ yr would produce only about 194 bits of information (Argyle 1977). This is far short of the $1.2 \times 10^5$ bits in a typical virus, and tiny compared to the $6 \times 10^6$ bits in a bacterium like *E. Coli*. The low gain in information

*I* from *N* trials is because the two things are related by $I = \log_2 N$. There are in principle two ways to circumvent this problem. One is that life in fact evolved solely on the Earth, but by some non-random, directed molecular process. The other is that life evolved on the Earth and other planets because they were seeded by biological molecules which *already had a large information content*. Both of these hypotheses have objections; but in view of the near-inevitability of this process shown above, the second appears to be the more plausible.

The idea that life on a given planet originates from the dead bits of other life also helps us understand the nature of the genetic code observed on Earth. Attempts at explaining the step from amino acids to simple organisms have a long history, but none has met with wide acceptance (Woese 1969, Nagyvary and Fendler 1974, Kuhn and Kuhn 1978, Dickerson 1978, Schopf 1978, Dose 1986, Sharov 2006). Natural processes do not account for the genomes of observed organisms, basically because of the slow accumulation of genetic information. By contrast, the hypothesis that some information was present at the outset, while it does not solve the problem, certainly alleviates it.

Traditional versions of panspermia are centred on the survivability of organic material exposed to a unique set of stresses, and especially on the stability of DNA. This is understandable, given the central role of this molecule in living organisms on present-day Earth; but caution is advisable about a preoccupation with this, as we will see below. The pioneering work of Lindahl (1993) focussed attention on the instability and decay of DNA in the water environment of modern living organisms. He noted that DNA is unstable to hydrolysis, oxidation and certain other processes, though damage can be countered by repair mechanisms as long as the material is in a normal environment at reasonable temperature and pressure. However, he noted that the rate of damage to DNA decreases with temperature, and left open the question of repair mechanisms that might operate in extreme environments. In fact, bacteria are known which thrive even in the cores of nuclear reactors, because they use an enzyme that can repair of order $10^4$ DNA strand breaks caused by ionizing radiation (Overmann, Cypionka and Pfenning 1992). As noted before, extensive studies have been done on the response of organic matter to electromagnetic radiation and ionizing particle radiation (Horneck 1981, 1993; Horneck et al. 1984, 2001; Dose and Klein 1996; Nicholson et al. 2000; Scalo and Wheeler 2002; Court et al. 2006). In the case of DNA, damage to its double helix commonly occurs in the form of single-strand breaks, double-strand breaks, interstrand cross-linking (particularly by proteins) and photochemically-produced lesions. Of these, the first is not normally lethal, but the others destroy the integrity of the molecule, effectively killing its host cell. On Earth, however, Nature has evolved a mechanism of protecting certain bacteria, namely via a dormant or spore mode. In a typical spore, the DNA in the nucleus is protected by a cortex and a tough outer coat which greatly extends survivability, allowing the organism to revivify when conditions become more congenial. Unprotected bacterial spores are killed by solar ultraviolet radiation in a few minutes, but ones shielded from direct rays can survive in the cold vacuum of space for several years, and survival times for spores embedded in stony material can be much longer. The timescale for inter-star panspermia, we recall from Section 2, is of order $10^5$ yr. It is therefore intriguing to note that recently reports have appeared of bacteria found in ancient Earthly formations, most notably a spore-forming species of *Bacillus* recovered from a 250 x $10^6$ yr old salt crystal (Vreeland, Rosenzweig and Powers 2000). Discoveries of this kind remain controversial, because of the possibility of contamination. Even a breath from the investigator may carry a single modern DNA molecule into an old sample, so that when amplified by the polymerase chain reaction currently used in laboratories, a false-positive detection of ancient life may result (Willerslev and Cooper 2005). However, a careful analysis of the stability of DNA against the production of double-strand breaks and the accumulation of single-strand breaks shows that the host spores may be expected to survive for periods of order $10^8$ yr, at least in regard to ionizing beta (electron) radiation at the intensities commonly found in rocks (Nicastro, Vreeland and Rosenzweig 2002). This does not alter the fact, emphasized above, that ultraviolet radiation *does* kill most organisms on the interstellar journeys of traditional panspermia. But it is important that geological sampling provides in principle a means of looking for fragments of old DNA and other organic debris, so allowing us to test models of panspermia.

Damage, repair and survival are central to any version of panspermia and the identification of the progenitors of modern organisms. DNA is not a likely progenitor, because it is more susceptible to radiation damage than other molecules. Also, there is no known way to generate life as we know it from a piece of DNA, partly because the instructions for the genetic code are not in the code itself, and damaged molecules do not have the transcription and translation abilities usually considered necessary for life. This problem may in principle be partly avoided by appeal to other properties of nucleotides or other molecules such as RNA (Trevors and Abel 2004, Warmflash and Weiss 2006).

However, it must be admitted that all versions of panspermia suffer from a hole in our knowledge, concerning how to go from an astrophysically-delivered entity which contains substantial information to one which has the characteristics of what we normally regard as life.

A possible candidate, which lies in the middle of this divide, is the virus. A typical virus contains of order $10^5$ bits of information, which is within an order of magnitude of that encoded in the simplest bacteria. The strongest argument against viruses as progenitors is that present-day ones need a host cell in which to multiply (see the Appendix). However, this argument is not totally convincing on closer examination, because it is known that conditions on the Earth were considerably different from those on the present planet, and have been more supportive of isolated viruses. That viruses may have pre-dated cells is indicated by the fact that several genes essential to the structure and replication of viruses are absent from the genomes of cells. These hallmark genes are present in many groups of DNA and RNA viruses, where they code for key proteins; and of course modern viruses are adept at changing their forms. It is not difficult to conceive of a primeval evolutionary path in which an astrophysically-delivered source gave rise on the Earth to RNA and then DNA viruses, which in turn evolved into eukaryotic cells. (The idea that viruses may be found in interstellar clouds is of course an old one, having been discussed by Hoyle and Wickramasinge 2000 and others.) Recent work on the genomics of viruses has led to a specific model of an ancient viral world (Koonin, Senkevich and Dolja 2006). It would be simplistic to apply this model directly to the primitive Earth, but at least it has the merit of being testable. Models for the primitive-Earth environment can be tested in the laboratory and in space. In both, care will be necessary to avoid contamination of any collected samples (see above). Meteorites from the Moon and Mars provide, of course, an obvious place to look for ancient organic remains, but are somewhat off our purpose because we are interested in material that has entered our solar system from outside. The physics of the panspermia process reduces the likelihood of finding genuine alien material in the inner solar system; and the near-Earth regime should be avoided because it is polluted, in part by human waste dumped from manned missions. For these reasons, the search for dust grains with the primordial imprint of proteinoids or pre-viruses should be made in the outer solar system.

Astrophysical data provide, in fact, the strongest constraints on possible progenitors for life (Section 2). In this regard, it is instructive to consider some numbers.

Sizes for the dust grains involved in panspermia are of order $10^{-5}$ cm, and surveys of the interstellar medium show that the material concerned is a mixture of silicates, graphite and water ice. Immediately, we encounter a physical constraint on the biological nature of the postulated organic matter attached to or embedded in the grain: the sizes of many spores (dormant cells) are of this order or larger. For example, *Bacillus subtilis* is an ovoid with a long axis typically in the 1-2 micron ($10^{-4}$ cm) range. This constraint can be circumvented, either by assuming that the panspermia objects are naked spores (which would be totally unprotected) or that they are nanobacteria (which are of order $10^{-7}$ cm and currently under intense study). However, we do not need to appeal to such alternatives, because as we saw above viruses (or their primitive precursors) are suitable candidates for panspermia. These have a wide range in size (Dimmock, Easton and Lep-pard 2007). The smallest, well-studied ones belong to the family *Parvoviridae* and are about 20 nm ($2 \times 10^{-6}$ cm), while the largest belong to the family *Poxviridae* and range up to 400 nm ($4 \times 10^{-5}$ cm). If we assume that a small protovirus had a size of 10 nm, then a dust grain of size $10^{-5}$ cm could accommodate roughly a thousand such. Of these, those near the centre would be better shielded than those near the surface. The latter would be in danger of having their information scrambled by ultraviolet photons and high-energy protons. In this connection, it should be noted that the large information content of many organisms, and especially cells, can be traced to the tight packing of their component proteins. The size constraint on the grains implies that any large molecules should maintain at least some of their packing, since an unfolded molecule will not fit into a typical grain. (The size limits of very small micro-organisms have been reported on by the Steering Group of the N.R.C. / N.A., 1999). The mechanism of protein folding, and other aspects of the behaviour of biological polymers, is not completely understood. It would be helpful to obtain more information on the behaviour of proteins under panspermia-like conditions. Such information would also help to establish the veracity of viral-world models for the early Earth. Our present knowledge indicates that the progenitors for panspermia were of virus size or smaller.

Timescales determine in large part whether organisms arrive on a planet like the Earth in a living or dead state. As we have seen, organisms in dust grains are killed by ultraviolet photons and/or cosmic ray particles on an astrophysically short timescale, while the grains travel from one star system to a nearby one on a timescale of $10^5$-$10^6$ yr, and the age of the Milky Way is of order $10^{10}$ yr. This implies that panspermia with liv-ing organisms is unlikely or rare, whereas panspermia with dead organisms is likely or common. In principle, therefore, our own solar system may harbour biological molecules that have come from numerous other places. This implies a kind of mongrel version of panspermia, where the genetic information in any given stellar system is derived from multiple progenitors. However, it must be admitted that the astrophysical data are uncertain, and can affect the argument in opposite directions. For example, if stars are born in denser-than-average clusters and organisms are transported in objects larger than grains, then the probability of 'living panspermia' is increased and so is the number of likely progenitors (Adams and Spergel 2005). Conversely, if life and especially highly-evolved intelligent life is sparse in the universe, then we necessarily fall back on 'dead panspermia', with the possibility of a few or even a unique progenitor (Wesson 1990). Irrespective of these uncertainties, it is apparent that the cosmic odds strongly favour the 'dead' option. How could this hypothesis be tested? One direct way is via new experiments on large biological molecules to improve our knowledge of how these fare when exposed to conditions like those expected during panspermia. Such experiments can be carried out with moderate cost, either on the ground or in space (Rabbow et al. 2005, Parnell, Mark and Brandstatter 2008). While we cannot obtain direct information on cosmic timescales, we can look into the behaviour of large organic molecules in oxidated or hydrolysed conditions over timescales longer than those previously considered. Another possible test involves chirality, the 'handedness' of biologically-important molecules. All known organisms on the Earth have the same chirality (Hartmann 1983, Panov 2005, Zagorski 2007). Why this should be so is not understood, and neither is it known if chirality is preserved during the extreme physical processes that accompany panspermia. But we expect that chirality will prove to be an important indicator about past panspermia, when in the future we are fortunate enough to obtain data on the life-forms of extrasolar planets.

In this section, the discussion has been of a general, information-based type; but astrophysical considerations mean that it is likely that organisms involved in panspermia arrive at a new planet in an inactivated or dead state. An appropriate name for this version of the theory is *necropanspermia*. While at first glance this may appear to be a contradiction in terms, we should recall that "sperm" means "seed". And there is fundamentally no reason why a broken and apparently dead organism should not give rise to a whole and living one.

3. Conclusion

Life is a process that generates information, whereas many other processes generate entropy (misinformation). Information provides an objective measure in assessing panspermia, circumventing the subjective issue of how to define 'living' versus 'dead', and suggests that particles like viruses (which have traditionally been viewed as occupying an intermediate position) may have been the progenitors of like.

Astrophysically, there is a certain inevitability about what might be termed the standard model of panspermia, in which particles of size $10^{-5}$ cm are accelerated by radiation pressure out of one planetary system, travel across interstellar space, and are decelerated in a reciprocal fashion when they enter another system. Biophysically, how-ever, this process is fraught with dangers. Most micro-organisms are inactivated or killed by ultraviolet light in the originating system, something which is augmented by short-wavelength radiation in interstellar space and the destination system. Cosmic rays play a relatively minor part, but add to the problem. Much of the biological damage involves strand breaks in large information-carrying molecules such as DNA and RNA. Some shielding is provided if the organisms are small enough to fit inside grains composed of silicates, carbon and ice. This reduces the exposure to photons, though is less effective for the particles of cosmic rays (including the low-energy ones of stellar winds). The problem can in principle be ameliorated by shielding the organisms in objects of larger size, but the flux of ejected matter falls with bigger carriers, and the astrophysical statistics favour smaller carriers. The conclusion is the one reached before by various workers (e.g. Secker, Lepock and Wesson 1994). Namely, that the vast majority of organisms reach a new home in the Milky Way in a technically dead state.

Resurrection may, however, be possible. Certain micro-organisms possess remarkably effective enzyme systems that can repair a multitude of strand breaks (e.g. Overmann, Cypionka and Pfenning 1992). This and other mechanisms for restarting life depend on having an hospitable environment at the new homesite. The version of the standard theory wherein life restarts from technically dead but information-carrying material may be dubbed necropanspermia. It avoids the problem that random chemical processes cannot account for the information encoded in the genomes of even simple Earthly organisms. Various tests of this and other versions of panspermia can be carried out: The direct collection of organic material in the outer solar system; the eventual de-termination of the chirality of lifeforms on extra-solar planets; the clarification of the evolution of the genomes of simple organisms and viruses on the Earth; and laboratory experimentation to see if genetic 'rubble' can reconstitute itself to form viable replicating molecules. The last option would involve an approach similar to the classic Miller-Urey experiments, and would be relatively cheap and straightforward to perform.

Panspermia has itself gone through many reincarnations, and it remains to be seen if the version of it suggested above will fare any better than its predecessors. More data are needed to see if the multiple iterations of panspermia will lead to an acceptable model.

Acknowledgements

This work grew out of earlier studies with J. Secker and J. Lepock. Grant support was partially supplied by N.S.E.R.C.

Appendix: Viruses and Panspermia

On the present Earth, viruses require the environments provided by host cells. This is the main reason why viruses are widely regarded as somehow secondary to cells. It is also the reason why the origin of viruses remains a controversial subject, a common opinion being that viruses originated in some way as pieces of cell material which became self-serving – rogue organisms in a world dominated by benevolent cells. In the present section, we wish to gather together various data on present-day viruses that counter the preceding opinion and point to a primitive, autonomous origin for viruses. There are a surprisingly large number of facts which are consistent with a primitive origin for viruses, possibly by panspermia. The following list is not exhaustive, and is not ordered in any systematic way; but its contents may be verified by reference to standard works and the modern literature on viruses. To the objective reasoner, it is clear that the common opinion of viruses being secondary to cells is suspiciously simplistic, because it neglects the enormous evolution and the great changes in physical conditions that follow from the billions of years of Earth's history.

For the same reason, it is impossible to be dogmatic about the origin of viruses. However, the following data are suggestive.

(a) Viruses are structurally distinct from cells. Viruses are assembled from pre-formed, independently-synthesized components; whereas cells arise from other, pre-existing cells. Both viruses and cells are genetically coded by RNA and DNA; but viruses contain only one type of nucleic acid, while cells contain both.

(b) It is conventional to discuss plant, animal and bacterial viruses separately, following the distinction between these categories shown by cells. But in fact the properties of the three classes of virus are very similar, as might be expected if viruses are more primitive than cells.

(c) Viruses have a high degree of adaptability to new environments. Some grow poorly when first isolated, but by the selection of mutants they adapt well when transferred from one culture to another. Virions or virus pieces naturally have a mutancy frequency of order 1 in $10^5$, and mutants which are not neutralized by molecular antibodies can carry forward a strain suited to a new environment.

(d)  Viruses have remarkable physical stability.  This is partly due to the mechanics of virus architecture, which produces regular, energetically-favourable structures; and partly to the inherent strength of their proteins, which while often irregular in shape are present in large numbers, so that a given virus may be more than 50% protein by weight.  The fact that viruses are assembled from subunits (see above) also leads to greater *genetic* stability, since a smaller unit lessens the likelihood of a disadvantageous mutation in the gene which specifies the unit.

(e)  The assembly of viruses from particles of protein is essentially a physical process.  Local conditions of temperature and pressure influence how often proteins are brought together by thermal movements, and so how a large number of weak bonds can lead to energetically-favourable structures.  Even in cases where other proteins are used during the construction of a virus, these are discarded as the virus attains its final form.  Thus, the forms of many viruses are similar to those of crystals, and in both cases the assembly process does not require specific genetic information.  (Some viruses are aided in assembly by genetics, however, while some are constructed directly from precursor proteins.)  Studies of the tobacco mosaic virus and others have shown that many viruses are created by a process which is effectively one of *self*-assembly, governed largely by physics.

(f)  The fact that some viruses infect their host cells appears to have led to the widespread view that viruses are interlopers in a cell-dominated world.  This view is subjective and misleading.  In some cases (e.g. certain bacteriophages) the DNA in the infected cell is carried there by the virus, which is therefore essential to some paths of evolution. Laboratory experiments have shown that when viral DNA is introduced to cellular DNA, the binding can take place at many different sites, and that the relative location of the two types is stable over hundreds of generations. Among the four possible permutations of single-stranded and double-stranded RNA and DNA, all four types of nucleic acid are found in viruses in bacteria, plants and animals (though with differing frequencies). That is, viruses and their genetic material are as ubiquitous as cells.

(g) There is a wide range in the sizes of viruses, in terms of physical diameter and molecular weight. (These are tabulated in standard sources, usually in nanometers for the principal dimensions and $M_r$ numbers for the DNA and RNA content, respectively.) The potato spindle tuber viroid is among the smallest, self-replicating pathogens known, and has only 359 nucleotides. (Viroids as opposed to viruses are small, circular structures of single-stranded RNA with 246 - 370 nucleotides, which do not code for proteins and replicate in the host cell nucleus.) Small fragments of DNA, with 100 - 140 nucleotides, are also important because their assembly into whole molecules is the preferred replication mode of many viruses, such as polyomavirus. Small viruses depend more on their host cells than do large ones, in the sense that large viruses can multiply when their host cells are not in the process of synthesizing DNA (the S phase). Most viruses have sizes in the range 20 - 300 nm and nucleic acid molecular weights of order $10^6$ - $10^8$. A typical dimension of 100 nm = $10^{-5}$ cm is the same order as the sizes of the dust grains implied by panspermia.

(h) The replication of genetic information in viruses is remarkably faithful. DNA replication happens with an error rate of only one mistake per $10^9$ - $10^{10}$ base-pair replica-tions, while RNA replication has an error rate of one per $10^3$ - $10^4$. Some viruses, such as polyomavirus, appear to use RNA as an aid to the replication of DNA. As noted above, the molecular measures of the nucleic acid in viruses cover a large range, and some genomes with $M_r$ = 1 - 2 x $10^8$ correspond to the material in 100 - 200 genes. The difference in the fidelity of DNA and RNA replication in present-day viruses has implications for their origin and evolution. Laboratory experiments show that incubation and dilution of certain types of virus RNA can cause the genetic information to be reduced to only 17% of the original sequence, indicating a kind of natural selection at the molecular level driven by physical factors. (The so-called defective-interfering viruses have defective RNA compared to their parents, and via their interactions with enzymes effectively create another replication / evolution route.) In the test-tube, bare RNA molecules can interact more freely than they do when in their parent viruses, and show a kind of evolutionary phase in response to environmental pressures that may have occurred on Earth before the appearance of cells. Hence the hypothesis of a primitive RNA world.

(i) The reason why some viruses kill the host cells on which they depend, sometimes leading to the demise of the human subject, is widely regarded as a mystery. The mechanism involved is not itself fully understood, though it may be that some viruses trigger self-programmed cell death (apoptosis). This would appear on the surface to be a case of a wrong path in evolution, since the death of the cell results usually in the death of the virus. However, this view is anthropomorphic. A different one is that viruses were present on the primitive Earth in some form, and have evolved to live an easy life in cells, while at the same time not 'caring' about their hosts. Then, the death of a number of in-dividual viruses would be of little importance; and even the extinction of all of the members of some type would be of small consequence, since its biological niche would presumably be taken over by some other type of virus. Even if many types of virus were to disappear due to the extirpation of their hosts, some types might survive by evolving back to their pre-cellular forms.

(j) The origin and evolution of viruses are two sides of the same problem. It exists because of lack of data, from laboratory experiments, fossil records in ancient rocks, and signs in extra-terrestrial material. In the absence of good data from these fields, it is not surprising that most workers continue to believe (by default) that viruses post-date cells. Specifically, viruses are widely believed to owe their parasitic dependence on cells to their derivation *from* cells. They could have arisen from devolved cells which lost the ability to function in a wider environment and become problematical parasites (rogue bacteria); or they could have come from fragments of cellular nucleic acid which abandoned their usual function for a dependent and destructive one (vagrant genes). Both versions of the cell-origin theory for viruses have grave objections, and the data we have listed show that viruses might have pre-dated cells or been coeval with them. Work to resolve this problem can usefully be done on several fronts and on specific topics. Laboratory studies would elucidate the evolution of viruses, particularly if it focussed on the difference in the error rate in replication of DNA and RNA (see above), which implies that an RNA virus can diversify its genetic code $10^5$ times faster than a DNA virus. Fossil studies would show directly what the virus / cell 'balance' was in the Precambrian era, which encompasses most of the Earth's geological history, provided a reliable marker can be found which differentiates between the two populations. Extra-terrestrial samples are routinely collected both on the Earth (meteorites) and in space (meteoritic dust), but again some distinctive marker is needed to identify the presence of viral material which is genuinely from space as opposed to originating from the Earth.

The problem of the origin and evolution of viruses is solvable. The import of the points (a) - (j) above is that at present we do not know if viruses and cells cohabited on the early Earth, or if one preceded the other. The answer to this question is obviously important for panspermia.